**Mathematical modeling of tumor therapy with oncolytic viruses:**
**Effects of parametric heterogeneity on cell dynamics**

Georgy P. Karev[1], Artem S. Novozhilov[1], Eugene V. Koonin[1,*]

[1]National Center for Biotechnology Information, National Library of Medicine, National Institutes of Health, Bethesda, MD 20894, USA.

*To whom correspondence should be addressed. Email koonin@ncbi.nlm.nih.gov

**Abstract**

**BACKGROUND:** One of the mechanisms that ensure cancer robustness is tumor heterogeneity, and its effects on tumor cells dynamics have to be taken into account when studying cancer progression. There is no unifying theoretical framework in mathematical modeling of carcinogenesis that would account for parametric heterogeneity.

**RESULTS:** Here we formulate a modeling approach that naturally takes stock of inherent cancer cell heterogeneity and illustrate it with a model of interaction between a tumor and an oncolytic virus. We show that several phenomena that are absent in homogeneous models, such as cancer recurrence, tumor dormancy, an others, appear in heterogeneous setting. We also demonstrate that, within the applied modeling framework, to overcome the adverse effect of tumor cell heterogeneity on cancer progression, a heterogeneous population of an oncolytic virus must be used. Heterogeneity in parameters of the model, such as tumor cell susceptibility to virus infection and virus replication rate, can lead to complex, time-dependent behaviors of the tumor. Thus, irregular, quasi-chaotic behavior of the tumor-virus system can be caused not only by random perturbations but also by the heterogeneity of the tumor and the virus.

**CONCLUSIONS:** The modeling approach described here reveals the importance of tumor cell and virus heterogeneity for the outcome of cancer therapy. It should be straightforward to apply these techniques to mathematical modeling of other types of anticancer therapy.

# BACKGROUND

Cancers are extremely complex systems that possess many features of robustness, i.e., they are systems that tend to maintain stable functioning despite various perturbations [1, 2]. The main two mechanisms that enable cancer robustness are functional redundancy, that comes from inherent heterogeneity of tumor cells, and feedback-control systems that facilitate survival of a tumor under adverse conditions, e.g., caused by anticancer drugs [3-6]. Many approaches to anticancer treatment have had limited success due to cancer robustness. An important challenge is to identify fragilities of cancers as robust biological systems and develop treatment strategies that take advantage of these weak links. Thus, a better understanding of the mechanisms that yield cancer robustness, particularly, functional redundancy, is vital.

Redundancy in cancers occurs at two levels. First, multiple copies of identical cells increase the likelihood of restoration of a tumor after treatment. Second, functional redundancy can be mediated by functionally equivalent but heterogeneous components (known as heterogeneous redundancy). Heterogeneous redundancy is thought to be central to tumor robustness. Although heterogeneity is a well-recognized characteristic of tumors that can lead to drug resistance, the current body of experimental and clinical data that relate to dynamic changes in intratumoral heterogeneity during progression, as well as to responses to various therapeutic strategies, is insufficient [2].

The term 'tumor heterogeneity' means the existence of distinct subpopulations of tumor cells with specific characteristics within a single neoplasm [7, 8]; in particular, surviving subpopulations of cells with metastatic potential after anticancer therapy can lead to tumor recurrence [9-11]. Tumor heterogeneity is a well documented phenomenon [3] as demonstrated by extensive cytogenetic analysis [12, 13]. Furthermore, many tumors are characterized by heterogeneous, often, non-random, spatio-temporal distribution of genetically heterogeneous tumor cells [4]. Distinct subpopulations of cells have been isolated from experimental and human neoplasms of every major histological type and location. These subpopulations are heterogeneous for many characteristics, such as morphology, growth rate, metastatic potential, karyotype, antigenicity, immunogenicity, biochemical properties, sensitivity to chemotherapeutic agents and radiation, etc. Genetic heterogeneity is, arguably, a major cause of acquired drug resistance of tumors [14, 15].

The dynamics of host-tumor system, which entails co-evolution under the selective pressure that is imposed by host environments, including antitumor drugs, is highly complex and nonlinear. Thus, to precisely define the conditions for successful therapy, mathematical models are needed. Extensive efforts have been dedicated over many years to mathematical modeling of cancer development and anticancer therapy. Stochastic models that take into account random mutations and cell proliferation proved to be useful in the context of epidemiology and statistical data [16], and for modeling cancer initiation and progression in terms of somatic evolution [17]. Deterministic models of tumor growth have proved valuable as well. Many of these have addressed avascular and vascular tumor growth taking advantage of methods borrowed from physics [18] but some use models from population biology to treat a tumor as a dynamic society of interacting cells [19-21]. A variety of mathematical approaches contribute to modeling cancer progression from different standpoints and take stock of various factors affecting tumor growth [22 and references therein, 23]. Combined analysis of tumor growth and anticancer therapies within the framework of mathematical modeling also produced a number of significant results [24, 25].

Inasmuch as tumor heterogeneity is one of the crucial factors in determining possible outcomes of anticancer treatment, it has to be incorporated and investigated in mathematical models. Although heterogeneity of tumor cell populations has been widely studied, there is no general approach that would provide a comprehensive description of intrinsic heterogeneity of tumors and would be amenable to qualitative and quantitative mathematical analysis. The existing mathematical models of anticancer therapy either do not include the effects of heterogeneity [e.g., 26, 27], or consider finite, usually, small, number of subpopulations, e.g., sensitive and resistant cells or proliferating and quiescent cells [e.g., 28, 29]. Other models take into account only spatial heterogeneity and do not address the important subject of genetic heterogeneity [e.g., 30, 31, 32], or use individual-based approach and simulation techniques and thus are, practically, not amenable to analytical mathematical analysis [e.g., 33, 34, 35].

The main goal of the present paper is to introduce a novel approach to model heterogeneity into the field of cancer modeling. We show that heterogeneous models, although still oversimplified, reflect qualitatively new phenomena, some of which are observed in experiments and clinical trials. We present an effective method to analyze these models and examine the predictions that such heterogeneous models can yield.

We illustrate our approach by addressing a complex process that involves both virus-cell interaction and tumor growth, namely, the interaction of the so-called oncolytic viruses with tumors. Oncolytic viruses are viruses that specifically infect and kill cancer cells but do not affect normal cells [36-39]. Many types of oncolytic viruses have been studied as candidate therapeutic agents, including adenoviruses, herpesviruses, poxviruses, reoviruses, paramyxoviruses, and retroviruses [37, 39]. Probably, the best-characterized oncolytic virus, that has drawn much attention, is ONYX-015, an attenuated adenovirus that selectively infects tumor cells with a defect in the *p53* gene [38, 40]. This virus has been shown to possess substantial antitumor activity and has proven relatively effective at reducing or eliminating tumors in clinical trials [41-43]. Although safety and efficacy remain major concerns, several other oncolytic viruses acting on different principles, including tumor-specific transcription of the viral genome, have been developed, and some of these viruses have entered or are about to enter clinical trials [37, 44-47]. Recently, synergistic use of immune therapy and oncolytic viruses has shown particular promise in cancer treatment [48].

The oncolytic effect can result from at least three distinct modes of virus-host interaction [37, 39]. The first mode involves repeated cycles of viral replication in the tumor cells leading to cell death and, consequently, to tumor reduction and, potentially, elimination. The second mode involves low-level virus reproduction that, however, results in the production of a cytotoxic protein that causes cell damage. The third mode consists in induction of antitumor immunity by virus infection of cancer cells. Cancer cells possess weak antigens for host immune sensitization. Virus infection causes inflammation and lymphocyte penetration into the tumor, with the virus antigens eliciting increased sensitivity to tumor necrosis factor-mediated killing.

Although the indirect modes of virus cancer therapy based on production of cytotoxic proteins or antitumor immunity might be promising, direct lysis of tumor cells by an oncolytic virus is the current mainstream strategy. Experiments on human tumor xenografts in nude mice have shown that the effect of oncolytic virus infection on tumors can range from no apparent effect, to reduction and stabilization of the tumor load (i.e., the overall size of a tumor), to elimination of the tumor [49]. Complete regression of tumors has been reported also in some patients treated with oncolytic viruses as part of clinical trials [50]. In a previous study [51], we presented a conceptual mathematical model of tumor cells-virus interaction which, depending on system parameter values, exhibits various behaviors including deterministic elimination of the cancer cells. Here, we further examine this model in conjunction with different possible heterogeneities of tumor cells, such as different susceptibility to infection, different death rates of infected cells, and others.

**RESULTS**

**Homogeneous mathematical models (phase-parameter portrait)**

In this section, we briefly present the main results of the analysis of a simple conceptual model of virus-tumor interaction [51]; these will be important for the further analysis described in the present work. The model allows for two populations of cells: uninfected tumor cells and infected tumor cells and is an extension of the previous work of Wodarz [52] to include an alternative non-linear functional response for infection rate.

The model, which considers two types of cells growing in the logistic fashion, has the following form:

$$\frac{dX}{dt} = r_1 X \left(1 - \frac{X+Y}{K}\right) - \frac{bXY}{X+Y},$$
$$\frac{dY}{dt} = r_2 Y \left(1 - \frac{X+Y}{K}\right) + \frac{bXY}{X+Y} - aY, \quad (1)$$

where $X$ is the size of the uninfected cell population; $Y$ is the size of the infected cell population; $r_1$ and $r_2$ are the maximum per capita growth rates of uninfected and infected cells respectively; $K$ is the carrying capacity; $b$ is the transmission coefficient (this parameter also includes the replication rate of the virus); and $a$ is the rate of infected cell killing by the virus (cytotoxicity). All the parameters of the model are supposed to be non-negative. Model (1) is subject to initial conditions $X(0) = X_0 > 0$ and $Y(0) = Y_0 > 0$. The concentration of viral particles is not explicitly included; it is assumed that virus abundance is proportional to infected cell abundance [53].

Rescaling model (1) by letting $X^*(t^*) = X(t)/K$, $Y^*(t^*) = Y(t)/K$, $t^* = r_1 t$ leads to the system

$$\frac{dX^*}{dt^*} = X^*(1-(X^*+Y^*)) - \frac{\beta X^* Y^*}{X^*+Y^*},$$

$$\frac{dY^*}{dt^*} = \gamma Y^*(1-(X^*+Y^*)) + \frac{\beta X^* Y^*}{X^*+Y^*} - \delta Y^*, \tag{2}$$

where $\beta = b/r_1$, $\gamma = r_2/r_1$, and $\delta = a/r_1$. In the following analysis, we suppress the asterisks to simplify the notations, but it should be clear that we use non-dimensional parameters and scaled sizes of cell populations, so that $X + Y \leq 1$ for any $t$.

The complete phase-parameter portrait of system (2) is shown in Fig. 1.

The model (2) exhibits all possible outcomes of oncolytic virus infection, i.e., no effect on the tumor (domains *I* and *II* in Fig. 1), stabilization or reduction of the tumor load (domains *IV* and *V*), and complete elimination of the tumor (domain *VIII*). Moreover there are two domains (domains *III* and *VII*) where the final outcome crucially depends on the initial conditions and can result either in failure of virus therapy or in stabilization (domain *III*) and elimination (domain *VII*) of the tumor.

**Heterogeneous model**

*General consideration*

The model (1) consists of two coupled, deterministic differential equations allowing for cell reproduction and death, and cell infection. This model is one of the mathematical models of tumor growth that treat a tumor as a dynamic society of interacting cells (e.g., [19, 20, 22]). Most population models suppose that all individuals have identical attributes, in particular, identical rates of growth, death and birth. This assumption simplifies computation, albeit at the cost of realism.

There are different ways to model heterogeneity. If it is assumed that heterogeneity of population is mediated by a structured variable (such as explicit space or age), we have to deal with partial differential equations (PDEs) (for application of space explicit models to tumor-virus interaction, see [54] and references therein; cell age is modeled, e.g., in [55]). By contrast, here, we explore heterogeneity of parameters (such as birth and death rates, and infection rate), i.e., we consider parameters as inherent and invariable properties of individuals, whereas parameter values can vary between individuals; any changes of mean, variance and other characteristics of the parameter distribution with time are caused only by variation of the population structure. This type of heterogeneity has been designated 'parametric' by Dushoff [56].

The most common method to take account of parametric heterogeneity is to divide a population into groups [28, 57, 58]. An important disadvantage of this approach is that heterogeneity within a group cannot be incorporated. Another approach is to consider the population as having a continuous distribution of parameters [59-62]. In the latter case, one usually has to deal with infinite-dimensional dynamical systems but, in a wide range of specific cases, there is an efficient analytical theory that allows one to reduce an initial distributed system to a system of ordinary differential equations (ODEs) [61, 62] (see also Mathematical Appendix). Here, we adopted this approach and emphasize that the applied reduction permits us to follow not only the total sizes of the populations, but also the time-dependent dynamics of the distributions, i.e., no information is lost when we switch from infinite-dimensional dynamical system to system of ODEs. It should be further emphasized that this reduction is not an approximation but results in an equivalent model, in contrast to the approaches commonly used to solve similar systems numerically [59, 63, 64].

The starting point of our analysis is the model (2) which is a non-dimensional version of the initial model (1). As a matter of fact, by assuming that one or more parameters are distributed through the populations of cells, we have to deal with the initial system (1) with dimensional parameters. However, the two parameters that are of particular interest are transmission coefficient $b$ and cytotoxicity $a$, and in the following, we assume that these parameters (or either of them) are distributed, whereas the net growth rates and carrying capacity are supposed to be the same for all tumor cells. Since non-dimensional parameters $\beta$ and $\delta$ differ from $b$ and $a$ only by linear scaling, the system with distributed non-dimensional parameters $\beta$ and $\delta$ can be analyzed without loss of generality.

*Distributed susceptibility*

Let us assume that transmission coefficient $\beta$ is distributed through the population of uninfected tumor cells. Denote $\mathrm{B}$ the set of possible values of $\beta$ and $x(t,\beta)$ the density of uninfected cells that have a given value of $\beta$ at moment $t$ (i.e., for any subset $\widetilde{\mathrm{B}} \subseteq \mathrm{B}$, the part of the population with trait values belonging to $\widetilde{\mathrm{B}}$ is given by $\int_{\widetilde{\mathrm{B}}} x(t,\beta)d\beta$, and the total size of the population, $X(t)$, is given by $\int_{\mathrm{B}} x(t,\beta)d\beta$).

The number of uninfected cells with a given value of $\beta$ that become infected per time unit is given by $\rho x(t,\beta)$, where $\rho = \rho(\beta, X, Y)$ is the rate of infection. We assume that this function depends only on the value of $\beta$ and the total sizes of the infected and uninfected cell subpopulations. This assumption reflects the fact that different cells can have different susceptibilities to virus infection, i.e., some of them can be infected with a relatively high probability whereas others are infected with a low or even zero probability; the latter case corresponds to the situation when a particular subpopulation of tumor cells cannot be reached by the virus.

Using the nonlinear transmission function in (2) we obtain that the change in subpopulation of uninfected cells with parameter value $\beta$ is given by $\beta x(t,\beta)Y(t)/(X(t)+Y(t))$. We also need to specify the law of growth of tumor cells with a fixed value of $\beta$. Let us assume that uninfected cells with a particular value of $\beta$ beget daughter cells with the same parameter value. For the net growth rate, we use the logistic law that depends on the total population sizes, $X$ and $Y$. We do not consider heterogeneity in the infected cell population, and, hence, the total amount of newly infected cells per time unit is given by $Y(t)\int_B \beta x(t,\beta)d\beta / (X(t)+Y(t))$, which should be taken into account when writing down the equation for the change in the infected cell population. By specifying initial conditions for $x(t,\beta)$ and $Y(t)$, we obtain the model in the form

$$\frac{\partial x(t,\beta)}{\partial t} = x(t,\beta)[1-(X(t)+Y(t))] - \frac{\beta x(t,\beta)Y(t)}{X(t)+Y(t)},$$

$$\frac{dY(t)}{dt} = \gamma Y(t)[1-(X(t)+Y(t))] + \frac{E_\beta(t)X(t)Y(t)}{X(t)+Y(t)} - \delta Y(t), \qquad (3)$$

where the following notation was used: $E_\beta(t) = \left(\int_B \beta x(t,\beta)d\beta\right)/X(t)$.

The initial conditions are

$$x(0,\beta) = x_0(\beta) = X_0 p(0,\beta), \quad Y(0) = Y_0. \qquad (4)$$

Here $x_0(\beta)$ is the initial distribution of $\beta$ in the population of uninfected cells, $X_0$ is the total number of uninfected cells at the initial moment, and $p(0,\beta)$ is the probability density function (pdf) of the initial distribution of $\beta$. Note that $E_\beta(t)$ can be rewritten in the form $E_\beta(t) = \int_B \beta p(t,\beta)d\beta$, where $p(t,\beta)$ is the pdf of the distribution of $\beta$ at time $t$, and, thus, $E_\beta(t)$ is the mean value of $\beta$ at moment $t$.

Integrating the first equation in (3) over $\beta$ we obtain the system

$$\frac{dX}{dt} = X(1-(X+Y)) - E_\beta(t)\frac{XY}{X+Y},$$

$$\frac{dY}{dt} = \gamma Y(1-(X+Y)) + E_\beta(t)\frac{XY}{X+Y} - \delta Y, \qquad (5)$$

where we suppressed the dependence of phase variables on $t$, and the initial conditions are $X(0) = X_0, Y(0) = Y_0$. In order to solve this system, we only need to know the explicit expression for $E_\beta(t)$. If we use the usual notation

$$M_\beta(\lambda) = \int_B e^{\beta\lambda} p(0,\beta) d\beta$$

for the moment generation function (mgf) of the initial pdf, it can be shown ([61, 62], see also Mathematical Appendix) that the current mean value of $\beta$ can be calculated using the mgf of the initial distribution of $\beta$; the exact expression is

$$E_\beta(t) = \frac{1}{M_\beta(q(t))} \left.\frac{dM_\beta(\lambda)}{d\lambda}\right|_{\lambda=q(t)},$$

where $q(t)$ is an auxiliary variable that satisfies the equation

$$\frac{dq}{dt} = -\frac{Y}{X+Y}, \quad q(0) = 0, \tag{6}$$

and $M_\beta(\lambda)$ is given as far as we specify the initial pdf $p(0,\beta)$.

In other words, model (3)-(4) is equivalent to the system of ODEs (5)-(6), which can be analyzed in the usual way; moreover, it is very easy to solve this system numerically (there are methods to obtain the solution of (3)-(4), e.g., [59], but these methods are usually computationally intensive and not so effective as numerical methods for ODEs).

Note that system (5) differs from (2) only by the time-dependent coefficient $E_\beta(t)$. In a number of cases, it is straightforward to determine the asymptotic behavior of $E_\beta(t)$ (i.e., when $t \to \infty$), and in this situation, we can use the theory of asymptotically autonomous differential equations to infer the final outcome of the cell dynamics using the parametric portrait (e.g., [65]). More importantly, however, the simultaneous knowledge of the time dependence of $E_\beta(t)$ and the parametric portrait of the homogeneous model (2) (Fig. 1) allows one to identify transient behavior of the solutions of (5) (and, accordingly, of model (3)), i.e., the behavior of solutions prior to the time moment when $E_\beta(t)$ becomes constant. This behavior can be of particular interest because this is one of the features that distinguishes model (3) from (2) and allows us to explore in detail the temporal dynamics of inhomogeneous biological systems.

Due to the non-negativity of $\beta$, we should consider only distributions with support on half-axis $\beta \geq 0$. To illustrate the possible dynamical behavior of the cell populations, we need the initial distribution of $\beta$, which is unknown. In the following, we assume that the initial pdf of $\beta$ is the gamma distribution on $[\eta, \infty)$ with positive parameters $k, s$ and $\eta \geq 0$. The choice of the gamma distribution can be justified by the fact that it can well approximate almost any unimodal distribution concentrated on the positive half-line. Other possible initial distributions include uniform, log-normal, Pareto, and many others; any of them can be incorporated in the model as far as there exist mgfs for the given $q(t)$ (we note, however, that, in some particular cases, mgf is not determined for any possible values of $q(t)$ (see Mathematical Appendix).

For the case of gamma distributed $\beta$, it is straightforward to show (see Mathematical Appendix) that $E_\beta(t) = \eta + k/(s - q(t))$, and $E_\beta(t) \to \eta$ when $t \to \infty$ (because $q(t) \to -\infty$, see (6)). However, in a particular situation, we might not reach the final point because $Y(t)$ can approach zero (see the equation for $q(t)$ in (6)).

In brief, we use the following approach to perform and present numerical simulations. First, we need to specify the initial conditions and the initial distribution of $\beta$, i.e., assign values of $\eta, k, s$. Here, for the sake of transparency, we have chosen to set the values of the mean $E_\beta(0) = \int_B \beta p(0, \beta) d\beta$ and variance $\sigma_\beta^2 = \int_B \beta^2 p(0, \beta) d\beta - E_\beta(0)^2$ of the initial distribution, as well as the bounds of the distribution if they are different from zero or infinity (for the gamma distribution, the bound that has to be specified is its left boundary, $\eta$). In the analyzed case (when we work with the distribution that has two free parameters, $k$ and $s$), this is sufficient to unambiguously determine the values of the parameters (obviously, however, this is not the case in the general situation).

It can be shown that, for model (3)-(4), the mean of the parameter distribution decreases monotonically with time ($dE_\beta(t)/dt < 0$ for any $t$) from the initial value $E_\beta(0)$ to the final value $\eta$. Loosely speaking, during the simulation, the parametric point with coordinates $(\gamma, \delta, E_\beta(t))$ travels in the parameter space of model (2); the movement occurs along the line connecting points $(\gamma, \delta, E_\beta(0))$ and $(\gamma, \delta, \eta)$ in the parameter portrait of (2) (see Fig. 1 for the cut of the parameter space for fixed $\gamma$). The speed of movement is determined by the parameters of the initial distribution, or, equivalently, by the initial mean and variance of the gamma distribution with the given left boundary.

Results of several numerical simulations of system (5)-(6) are shown in Fig. 2. The parameter values were chosen such that we start in domain *VIII* (Fig. 1) (eradication of the tumor), cross domain *VII* (bistable situation), and end up in domain *I* (no effect of virus therapy). The solutions shown in Fig. 2 reflect the fact that the degree of heterogeneity plays an important role in the model dynamics. The parameter values and initial conditions are the same for all four simulations; the difference comes from different initial variances of $\beta$; the greater the initial variance the faster we reach the unfavorable domain *I*. Conversely, the initial variance of the distribution can be small enough such that the time during which the size of the tumor remains negligible ($X + Y$ is close to zero) is comparable with the life-time of a patient; this emphasizes that we need to know not only the final state of $E_\beta(t)$ but also its transient behavior.

Fig. 2 shows the phenomenon of tumor recurrence after a relatively long period of small tumor load, a phenomenon that is not seen in the original homogeneous model (2). Clearly, reappearance of the tumor is a probable outcome of tumor-specific virus treatment if there are cancer cells that are inaccessible to the virus (or are accessible at an extremely slow rate compared to the characteristic time of virus propagation): after the virus kills off the susceptible tumor cells and is cleared (given that it cannot infect healthy tissues), these resistant cancer cells can develop into a new tumor. It should be emphasized that, in the simulations in Fig. 2, all the cells have non-zero probability to be infected ($\eta > 0$), but the final outcome is tumor recurrence, i.e., the tumor dynamics is strongly affected by cell heterogeneity. This example shows that heterogeneous model (3), together with dynamical regimes inherited from model (2), possesses new regimes, and, thus, is more general.

The change of the mean parameter value for the cases presented in Fig. 2 is shown in Fig. 3. It can be seen that higher rate of change corresponds to greater values of the initial variance; in addition, in none of the analyzed cases $E_\beta(t)$ reaches the final value $\eta$, because the speed of movement is determined by two simultaneously acting factors: the characteristics of the mgf and the equation for the auxiliary variable $q(t)$ for which $dq(t)/dt \approx 0$ starting from some time $t$.

Other possible cases can be similarly analyzed; for example, if we choose parameters such that the starting point belongs to domain *IV* (Fig. 1a) and the final point belongs to domain *II*, it is not difficult to predict that, first, there will be a short period of time when the tumor load remains constant, then the tumor starts growing linearly (when $E_\beta(t)$ belongs to domain *V*), and in domain *II*, the tumor grows under the logistic law (Fig. 4). Thus, accommodation of heterogeneity results in model dynamics that reflects the phenomenon of temporary dormancy of the tumor (Fig. 4).

*Distributed susceptibility and distributed cytotoxicity*

Together with differential susceptibility considered in the previous section, parameter $\delta$ (the rate of killing of infected cells) also may be assumed to take different values in the infected cell population. The population of infected cells can be heterogeneous on its own such that some cells are more likely to die faster than others when infected by the virus. From different perspective, we can attribute this heterogeneity to the virus, i.e., the virus population is assumed to consist of strains that differ in their ability to kill cells. In previous work [51, 52], it has been shown that, in some parameter domains, there is an optimal level of cytotoxicity that is necessary to maximally reduce the tumor size if other parameter values are fixed. In addition, it is sometimes desirable to kill as many cells as possible during a short period of time; such situation would favor using different viruses for therapeutic purposes. Moreover, the latter assumption on heterogeneous virus can help one to interpret the model equations which, assuming that $\delta$ takes values from set $\Delta$, and $y(t,\delta)$ is the density of infected cells having parameter value $\delta$, take the form:

$$\frac{\partial x(t,\beta)}{\partial t} = x(t,\beta)[1-(X(t)+Y(t))] - \frac{\beta x(t,\beta)Y(t)}{X(t)+Y(t)},$$
$$\frac{\partial y(t,\delta)}{\partial t} = \gamma y(t,\delta)[1-(X(t)+Y(t))] + \frac{E_\beta(t)X(t)y(t,\delta)}{X(t)+Y(t)} - \delta y(t,\delta), \quad (7)$$

where $E_\beta(t)$ is defined as above, and the initial conditions are

$$x(0,\beta) = x_0(\beta) = X_0 p_1(0,\beta), \quad y(0,\delta) = y_0(\delta) = Y_0 p_2(0,\delta). \quad (8)$$

Note that in (7) both cell populations are closed under reproduction; that is, e.g., the uninfected cell that has parameter value $\beta^*$ can produce daughter cells only with the same parameter value (the same holds for the infected cell population). At the same time, it follows from (7) that, if an uninfected cell was infected by the virus from an infected cell with the parameter value $\delta^*$, it falls into the same class. If we assume that $\delta$ is an attribute of the cell population, this assumption is difficult to justify, whereas if $\delta$ is an attribute of the virus, the assumption follows from obvious considerations.

System (7)-(8) is equivalent to the following system of ODEs (see Mathematical Appendix):

$$\frac{dX}{dt} = X(1-(X+Y)) - E_\beta(t)\frac{XY}{X+Y},$$
$$\frac{dY}{dt} = \gamma Y(1-(X+Y)) + E_\beta(t)\frac{XY}{X+Y} - E_\delta(t)Y, \quad (9)$$
$$X(0) = X_0, \quad Y(0) = Y_0$$

where the mean parameter values are

$$E_\beta(t) = \frac{1}{M_\beta(q_1(t))} \frac{dM_\beta(\lambda)}{d\lambda}\bigg|_{\lambda=q_1(t)}, \quad E_\delta(t) = \frac{1}{M_\delta(q_2(t))} \frac{dM_\delta(\lambda)}{d\lambda}\bigg|_{\lambda=q_2(t)},$$

for the given mgf of $p_1(0,\beta)$ and $p_2(0,\delta)$, and the auxiliary variables can be found from

$$\frac{dq_1}{dt} = -\frac{Y}{X+Y}, \quad \frac{dq_2}{dt} = -1, \quad q_1(0) = 0, q_2(0) = 0. \tag{10}$$

To analyze system (9)-(10), one has to specify initial conditions and initial distributions for parameters $\beta$ and $\delta$. As before, to obtain some insight into the possible dynamical behavior of (9), we can consider the parameter portrait (Fig. 1) and time-dependent changes of the mean parameter values. In the case under consideration $E_\beta(t)$, as before, moves from top to bottom in Fig. 1, whereas $E_\delta(t)$ goes from right to left. Note that the speed of movement in the $\delta$-direction is determined only by the properties of the mgf, because $q_2(t) = -t$ from (10) and $q_2(t)$ does not depend on the population sizes.

Generally, the transient behavior of cell populations (prior to reaching the final state – if it is ever reached) can be quite complex. This can be seen in Fig. 4 where the solutions of (9)-(10) are shown (Fig. 5a), and the path of the mean parameter values depicted (Fig. 5b). Due to the structure of the parametric portrait of (2), the mean parameter values can pass through domains of qualitatively different behavior many times, thus resulting in complex and unpredictable picture. Moreover, even if the initial and final parameter points belong to the same domain, during transient process, the mean parameter values can visit other domains (Fig. 6).

In Fig. 6, the initial and final parameter values belong to the most favorable domain *VIII* (Fig. 1), in which complete eradication of the tumor cells occurs. On its way to extinction, however, the tumor load behaves in an irregular, complex way. This example shows the possibility that complex and erratic observational data can be explained not only by random effects and noise but also by the innate heterogeneity of the cell and virus populations.

In general, both distributed susceptibility of uninfected cells and distributed cytotoxicity of the virus do not favor the tumor treatment and decrease chances for effective cure by increasing tumor robustness; one of the possible ways to overcome this problem (within the framework of the considered models) is suggested in the next section.

*Distributed susceptibility and distributed virulence*

The inherent heterogeneity of tumor cells implies that cells differ in their susceptibility to virus infection, and this situation was modeled in two previous sections. In general, it can be assumed that the virus population is also heterogeneous in the sense that it contains virus strains with different ability to infect tumor cells. Inasmuch as model (1) does not explicitly model the virus dynamics, we can incorporate the effect of virus heterogeneity into the infected cell population, and, hence, consider the transmission coefficient $\beta$ as the product of the susceptibility of uninfected cells $\beta_1$ and the virus replication rate $\beta_2$.

As before, denote $B_1$ the set of possible values of $\beta_1$ and $x(t,\beta_1)$ the density of uninfected cells having the given value of $\beta_1$ at the moment $t$; similarly, denote $B_2$ the set of possible values of $\beta_2$ and $y(t,\beta_2)$ the density of uninfected cells with the given value of $\beta_2$ at the time moment $t$. The number of cells with susceptibility $\beta_1$ newly infected by the virus produced from previously infected cells with replication rate $\beta_2$ is given by $\beta_1\beta_2 x(t,\beta_1)y(t,\beta_2)/(X(t)+Y(t))$. The total number of cells that leave the uninfected population and have susceptibility $\beta_1$ is $\beta_1 x(t,\beta_1)\int_{B_2}\beta_2 y(t,\beta_2)d\beta_2/(X(t)+Y(t))$; and the total change in the infected cell population with virus replication rate $\beta_2$ due to the infection process is given by $\beta_2 y(t,\beta_2)\int_{B_1}\beta_1 x(t,\beta_1)d\beta_1/(X(t)+Y(t))$. Combining the above assumptions, we obtain the model

$$\frac{\partial x(t,\beta_1)}{\partial t} = x(t,\beta_1)[1-(X(t)+Y(t))] - \frac{\beta_1 x(t,\beta_1)E_{\beta_2}(t)Y(t)}{X(t)+Y(t)},$$

$$\frac{\partial y(t,\beta_2)}{\partial t} = \gamma y(t,\beta_2)[1-(X(t)+Y(t))] + \frac{\beta_2 y(t,\beta_2)E_{\beta_1}(t)X(t)}{X(t)+Y(t)} - \delta y(t,\beta_2),$$

(11)

where the notations $E_{\beta_1}(t) = \left(\int_{B_1} \beta_1 x(t,\beta_1)d\beta_1\right)\!/X(t)$, $E_{\beta_2}(t) = \left(\int_{B_2} \beta_2 y(t,\beta_2)d\beta_2\right)\!/Y(t)$ were used for the current mean values of the corresponding parameters, and the initial conditions are

$$x(0,\beta_1) = x_0(\beta_1) = X_0 p_1(0,\beta_1),\ y(0,\beta_2) = y_0(\beta_2) = Y_0 p_2(0,\beta_2),\quad (12)$$

The model (11)-(12) is equivalent to the system of ODEs

$$\frac{dX}{dt} = X(1-(X+Y)) - E(t)\frac{XY}{X+Y},$$
$$\frac{dY}{dt} = \gamma Y(1-(X+Y)) + E(t)\frac{XY}{X+Y} - \delta Y,\quad (13)$$
$$X(0) = X_0,\ Y(0) = Y_0$$

where the mean parameter value is $E(t) = E_{\beta_1}(t)E_{\beta_2}(t)$, and

$$E_{\beta_1}(t) = \frac{1}{M_{\beta_1}(q_1(t))}\left.\frac{dM_{\beta_1}(\lambda)}{d\lambda}\right|_{\lambda=q_1(t)},\quad E_{\beta_2}(t) = \frac{1}{M_{\beta_2}(q_2(t))}\left.\frac{dM_{\beta_2}(\lambda)}{d\lambda}\right|_{\lambda=q_2(t)}$$

are expressed with the mgfs of $p_1(0,\beta_1), p_2(0,\beta_2)$. The auxiliary variables can be found from the equations

$$\frac{dq_1}{dt} = -E_{\beta_2}(t)\frac{Y}{X+Y},\quad \frac{dq_2}{dt} = E_{\beta_1}(t)\frac{X}{X+Y},$$
$$q_1(0) = 0,\ q_2(0) = 0$$

(14)

As before, we have to specify the initial pdfs for $\beta_1$ and $\beta_2$. Now we have $dq_2(t)/dt \geq 0$, and it can be shown that $dE_{\beta_2}(t)/dt \geq 0$, that is, formally, the movement in the $\beta_2$-direction goes from smaller to greater mean values. In order not to deal with an infinitely large coefficient, it is reasonable to assume that the initial distribution of $\beta_2$ is determined on a closed interval $[c_1, c_2]$. In this case, we can choose as the initial distribution the beta distribution on $[c_1, c_2]$.

To get insight into the transient behavior of the model solutions, we have to consider the product of the mean parameter values. Again, we move along the line in the $\beta$-direction (Fig. 1), starting from $E_{\beta_1}(0)E_{\beta_2}(0)$, with the asymptotic state $\eta c_2$. The important difference now is that the function $E(t)$ does not have to be monotonic (Fig. 7).

In Fig. 7, the results of numerical simulation of system (13)-(14) are shown together with functions $E_{\beta_1}(t)$, $E_{\beta_2}(t)$ and $E(t) = E_{\beta_1}(t)E_{\beta_2}(t)$. We start in domain *VIII* (elliptic sector in Fig. 1), and the asymptotic state is in domain *VII* where two opposite outcomes are possible, namely, complete tumor eradication or logistic tumor growth. The initial conditions and parameter values are the same for both cases; the two cases differ only in the initial variance of the $\beta_2$ distribution. It can be seen from Fig. 7 that even a small difference in the variance of $\beta_2$ may yield dramatically different results: in the first case, the tumor is cured, whereas, in the second case, virus therapy fails. This example emphasizes that, to predict the outcome of oncolytic virus therapy, it is necessary to know not only the initial sizes of the cell populations but also the degree of heterogeneity of all the parameters under consideration, in particular, the virus replication rate.

**Discussion and conclusions**

We presented a general framework within which to construct and analyze mathematical models of anticancer treatment, with a special emphasis on tumor heterogeneity. Conceptually, this approach is connected to the theoretical considerations of Kitano on cancer robustness [1, 2]. At this initial step of analysis, it appears most important that the mathematical models reproduce important behaviors of tumors at the qualitative level. Using the previously developed model of tumor cell-oncolytic virus interaction [51], we show that qualitatively distinct behaviors that are absent in the homogeneous model appear as natural consequences of tumor and oncolytic virus heterogeneity. For example, our model accounts for cancer recurrence, and the time until reappearance of the tumor crucially depends on the heterogeneity of the transmission coefficient (Fig. 2). It should be emphasized that the effects of the tumor cell heterogeneity are not limited to trivial resistance of a subpopulation of cells to the virus, i.e., recurrence might ensue even when all cells have a non-zero probability to be infected. Another phenomenon missing in the homogeneous model is constant tumor size during a particular period of time (Fig. 4). Heterogeneity in parameters can lead to complex, time-dependent behaviors of the tumor (Fig. 5 and 6). Thus, interpretation of the results of anticancer therapy should take into account the possibility that irregular, quasi-chaotic behavior can be caused not only by random fluctuations but also by the heterogeneity of the tumor and the virus (Fig. 6). All the simulations presented here reveal the effect of the level of heterogeneity on tumor dynamics. The most obvious case is shown in Fig. 7 where different initial variances of parameter distribution lead either to tumor eradication or to logistic tumor growth, i.e., failure of virus therapy.

Analysis of the models proves that tumor heterogeneity increases cancer robustness, in agreement with the theoretical considerations of Kitano [2]. The results presented here further show that, to counter this adverse effect of tumor heterogeneity, it should be possible to employ a heterogeneous population of an oncolytic virus (see model (11) and Figure 7). The interaction of the two non-homogeneous populations, the tumor and the oncolytic virus, may result in complete elimination of the tumor.

We applied a previously developed, general mathematical technique [61, 62] to investigate non-homogeneous models of complex tumor cells-virus interaction. The advantages of this modeling approach are as follows. This approach can account for different types of parametric heterogeneity of the analyzed populations. To infer the consequences of heterogeneity, we use well-known mathematical tools, such as bifurcation analysis, which identifies points of qualitative change in system dynamics. The theory of heterogeneous populations [61, 62, 66] allows one to reduce the models to systems of ODEs that, in many cases, can be explored analytically or, if this is not possible, can be solved numerically with high precision.

Cancer is an evolving system, and the main evolutionary forces are selection, mutation and random drift. It should be noticed that our approach explicitly examines only selection. The techniques is applied to deterministic systems, and it is our belief that such system are important for modeling purposes, although analysis of extinction phenomena, such as tumor eradication, may require stochastic factors to be considered. The principal obstacle to the validation of these models against empirical data and their use for prediction purposes is the necessity to know the initial distribution of the model parameters. It should be emphasized that knowledge of mean, variance, and any finite number of the moments of the distribution is not sufficient to describe the evolution of the system over indefinite time [62]. The entire distributions have to be known. However, there is a way to bypass this problem. First, knowledge of several first moments is sufficient to estimate tumor evolution over short time spans. Second, to model the evolutionary process, we need to know how the mean parameter values behave with time. Using the theory of heterogeneous populations, we can infer generic properties of these functions and identify them from empirical data [67].

To conclude, the approach developed here allows one to take stock of such a complex aspect of cancer as tumor heterogeneity and apply effective analytical techniques to the analysis of heterogeneous models of tumor evolution. Although, in this study, we analyzed the specific case of a tumor interaction with an oncolytic virus, it is our hope that these techniques will prove useful in other systems that include interaction between tumor cells and anticancer agents.

**Fig. 1.** Phase-parameter portrait of system (2) given as a cut of the positive parameter space $(\gamma, \beta, \delta)$ for an arbitrary fixed value of $0 < \gamma < 1$ (a) and $1 < \gamma$ (b). The boundaries of domains are $\alpha_1 = \{(\delta, \beta, \gamma): \delta - \beta = 0\}$, $\alpha_2 = \{(\delta, \beta, \gamma): \gamma - \delta = 0\}$, $\alpha_3 = \{(\delta, \beta, \gamma): \gamma\beta - \delta = 0\}$, and $\alpha_4 = \{(\delta, \beta, \gamma): \gamma - \delta - 1 + \beta = 0\}$.

**Fig. 2.** Solutions of system (5)-(6) with gamma distributed parameter $\beta$ on $[1.5, \infty)$. Uninfected cells, $X(t)$, infected cells, $Y(t)$, and the total tumor load, $X(t) + Y(t)$, are shown in blue, green and black, respectively. The initial conditions $X(0) = 0.5$, $Y(0) = 0.1$, parameter values $\gamma = 1$, $\delta = 2$. The initial mean of distribution $E_\beta(0) = 2.5$, the initial variances 0.06 (a), 0.1(b), 0.3(c), 0.4(d).

**Fig. 3.** $E_\beta(t)$ versus time for the cases presented in Fig. 2. The curves from top to bottom correspond to panels (a)-(d) in Fig. 2.

**Fig. 4.** Solutions of system (5)-(6) with gamma distributed parameter $\beta$ on $[0.1, \infty)$. Uninfected cells, $X(t)$, infected cells, $Y(t)$, and the total tumor load, $X(t) + Y(t)$, are shown in blue, green, and black, respectively. The initial conditions are $X(0) = 0.5$, $Y(0) = 0.1$, parameter values $\gamma = 0.5$, $\delta = 0.3$. The initial mean of distribution $E_\beta(0) = 3.5$, the initial variances 1.5 (a) and 0.5(b).

**Fig. 5.** (a) Solutions of system (9)-(10) with gamma distributed parameters $\beta$ on $[1, \infty)$ and $\delta$ on $[1, \infty)$. Uninfected cells, $X(t)$, infected cells, $Y(t)$, and the total tumor load, $X(t) + Y(t)$, are shown in blue, green and black, respectively. The initial means of the distributions are $E_\beta(0) = 11$ and $E_\delta(0) = 10$, initial variances are $\sigma_\beta^2(0) = 8$ and $\sigma_\delta^2(0) = 0.5$. The initial conditions are $X(0) = 0.5$, $Y(0) = 0.1$, and $\gamma = 1$. (b) The parametric curve $(E_\beta(t), E_\delta(t))$ in the parameter space (compare with Fig. 1b)

**Fig. 6.** (a) Solutions of system (9)-(10) with gamma-distributed parameters $\beta$ on $[2, \infty)$ and $\delta$ on $[0.9, \infty)$. Uninfected cells, $X(t)$, infected cells, $Y(t)$, and the total tumor load, $X(t) + Y(t)$, are shown in blue, green and black, respectively. The initial means of distributions are $E_\beta(0) = 11$ and $E_\delta(0) = 10$, initial variances are $\sigma_\beta^2(0) = 8$ and $\sigma_\delta^2(0) = 0.5$. The initial conditions $X(0) = 0.5$, $Y(0) = 0.1$ and $\gamma = 0.7$. (b) The parametric curve $(E_\beta(t), E_\delta(t))$ in the parameter space

**Fig. 7.** Solutions of system (13)-(14) with both uninfected cell specific and infected cell specific distributions of transmission coefficient $\beta$. $\beta_1$ is gamma-distributed on $[0.6, \infty)$ and $\beta_2$ is beta-distributed on $[0, 2.5]$. Uninfected cells, $X(t)$, infected cells, $Y(t)$, and the total tumor load, $X(t) + Y(t)$, are shown in blue, green and black, respectively. The initial means of distributions are $E_{\beta_1}(0) = 2.9$ and $E_{\beta_2}(0) = 0.6$, initial variances are $\sigma^2_{\beta_1}(0) = 1.9$ and $\sigma^2_{\beta_2}(0) = 0.12$ (a) $\sigma^2_{\beta_2}(0) = 0.11$ (b). In panels (c) and (d), the mean parameter values $E_{\beta_1}(t)$, $E_{\beta_2}(t)$ and $E(t) = E_{\beta_1}(t)E_{\beta_2}(t)$ are shown for cases (a) and (b), respectively

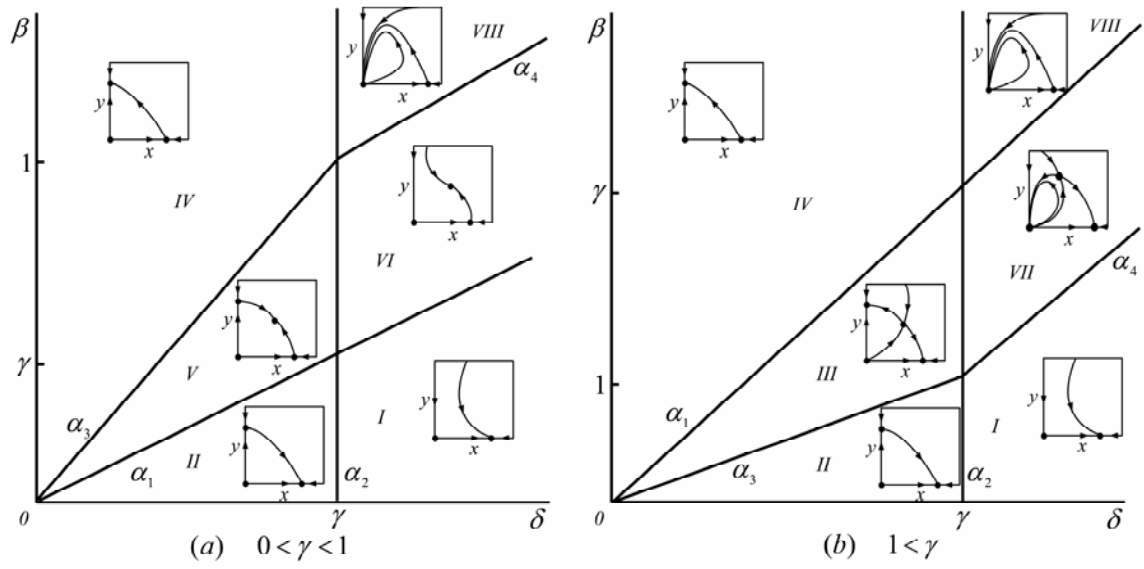

**Fig. 1**

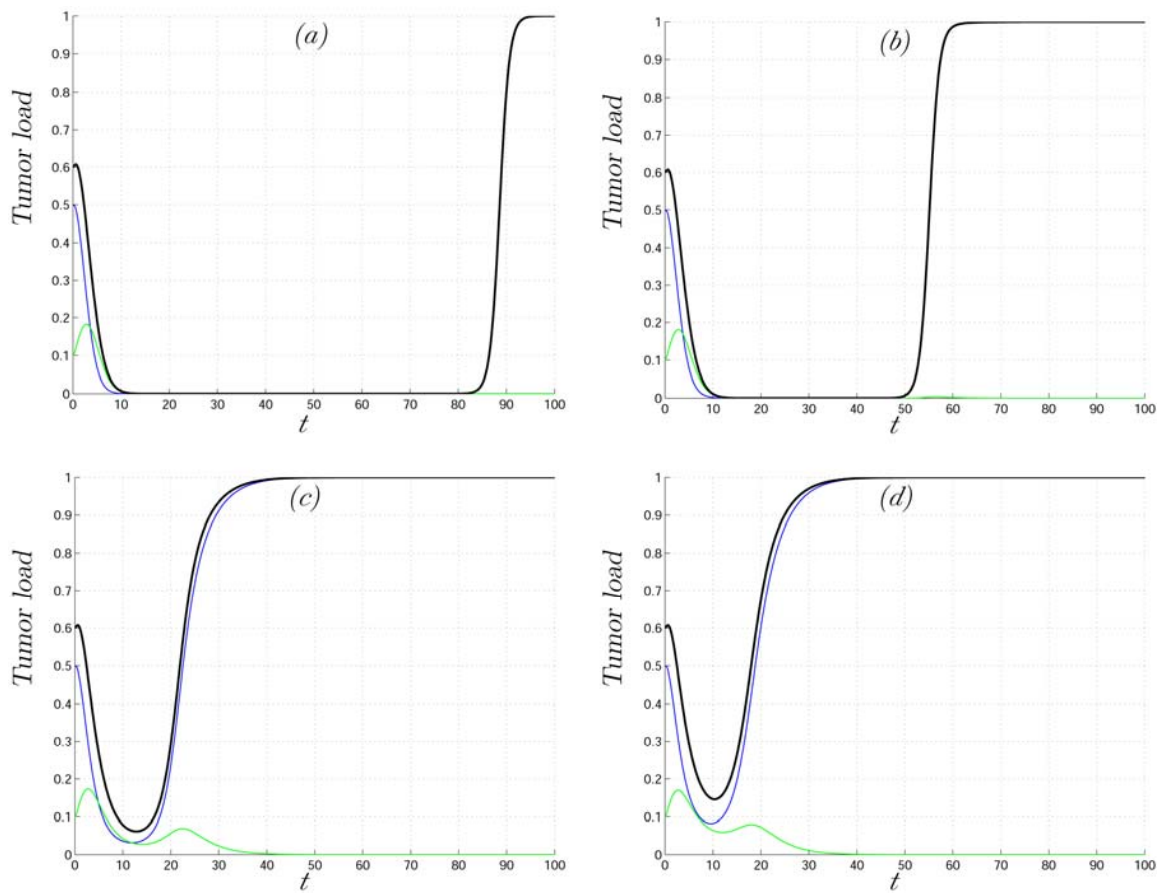

**Fig. 2**

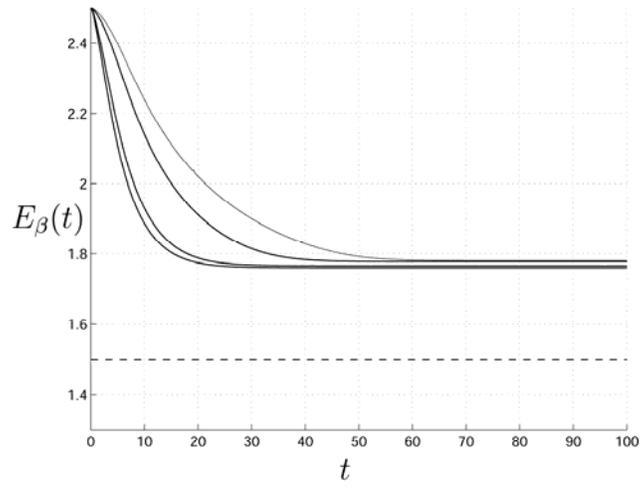

**Fig. 3**

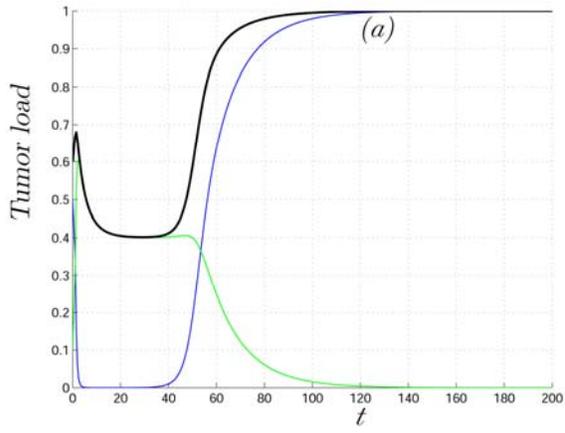 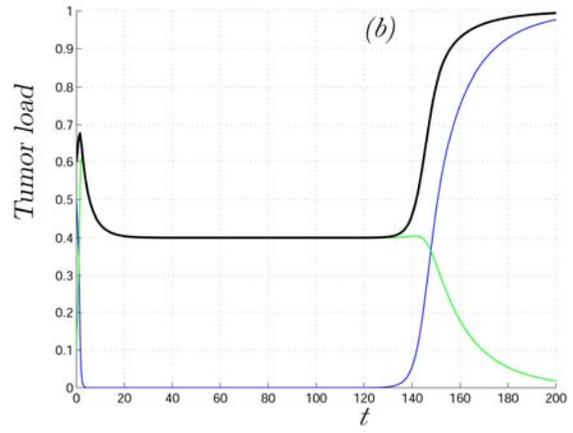

**Fig. 4**

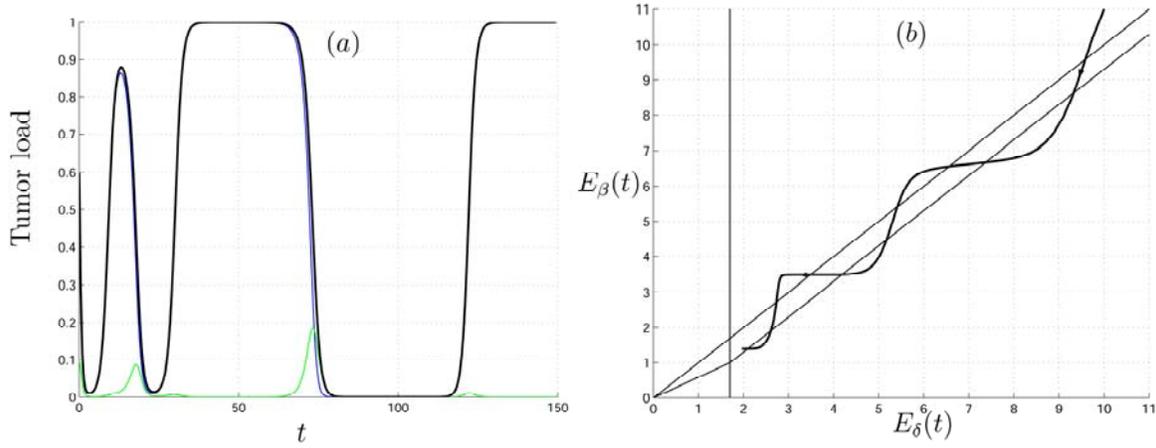

**Fig. 5**

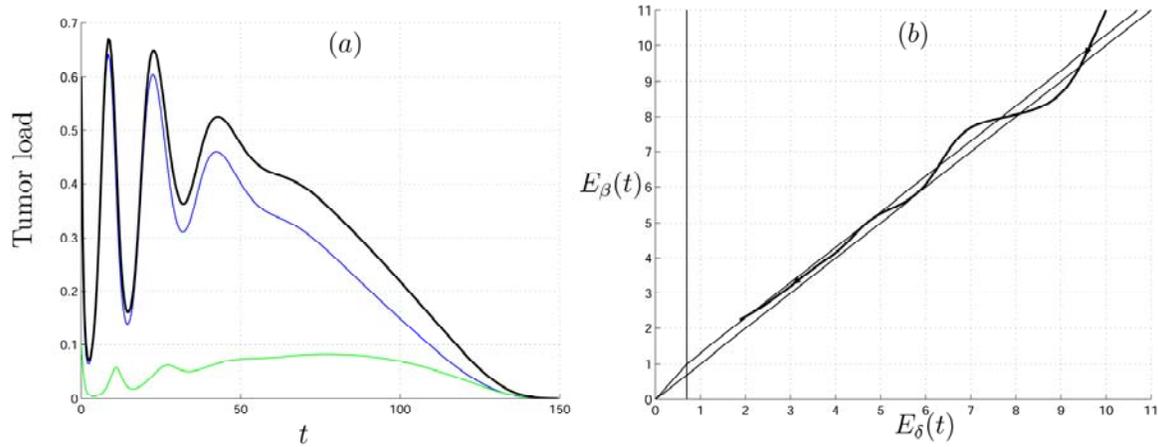

**Fig. 6**

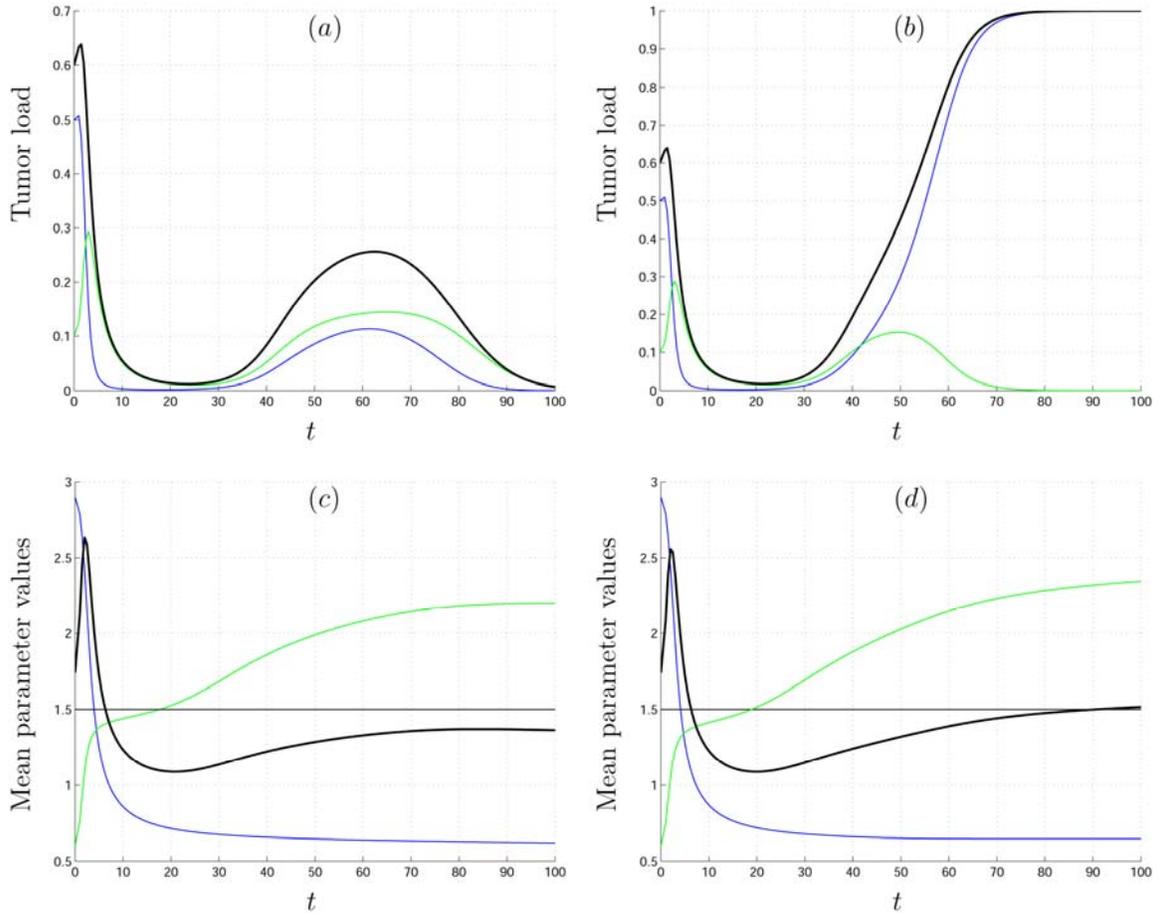

**Fig. 7**

**Mathematical Appendix**

**On Mathematical Theory of Inhomogeneous Community Models**

Let us consider a model of a community consisting of $m$ interacting populations in the general form

$$dN^j/dt = N^j F^j(\mathbf{N},\mathbf{a}), \quad j=1,\ldots m,$$

where $N^j$ is the size of $j$-th population, $\mathbf{N}=(N^1, \ldots, N^m)$. The growth rate $F^j$ can depend on parameters and $\mathbf{a}=(a_1, a_2, \ldots a_s)$ is the vector-parameter with the domain $A$. We suppose here that every individual is characterized by its own value of vector-parameters $\mathbf{a}$. Let $l^j(t,\mathbf{a})$ be the density of the $j$-th population along the vector-parameter $\mathbf{a}$ at time $t$. In [1-4] the following model of an inhomogeneous community was explored:

$$dl^j(t,\mathbf{a})/dt = F^j(\mathbf{N},\mathbf{a})\, l^j(t,\mathbf{a}), \tag{A.1}$$

$$N^j(t) = \int_A l^j(t,\mathbf{a})\, d\mathbf{a}.$$

We suppose here that the fitness of individuals in the $j$-th population, defined by the formula $F^j(\mathbf{N},\mathbf{a})=[dl^j(t,\mathbf{a})/dt]/l^j(t,\mathbf{a})$, depends on the parameter $\mathbf{a}$ and on total sizes of all subpopulations but not on the parameter distribution inside the subpopulations.

This type of models can well describe, e. g., cooperative dynamics of populations or resource-consumer model of inhomogeneous community, but are not applicable to prey-predator or epidemiological problems, where the interactions between populations should be "balanced" (e.g., the number of newly infected individuals should be equal to decrease of the number of susceptible individuals). Nevertheless, the main ideas and techniques used in [1] can be applied to "balanced" inhomogeneous community models with slight modifications.

We do not develop here a general theory of such models; instead, we consider a model of a biological community composed of two interacting populations depending on parameters:

$$\frac{\partial x(t,a)}{\partial t} = x(t,a)[F_1(X,Y,E_b(t)) + aG_1(X,Y,E_b(t))],$$
$$\frac{\partial y(t,b)}{\partial t} = y(t,b)[F_2(X,Y,E_a(t)) + bG_2(X,Y,E_a(t))],$$
(A.2)

assuming that parameters $a$ and $b$ are distributed. Here $x(t,a)$, $y(t,b)$ are the population densities, $X(t) = \int_A x(t,a)da$, $Y(t) = \int_B y(t,b)db$ are the total sizes of the cell populations. Denote $p_1(t;a)=x(t,a)/X(t)$, $p_2(t;b)=y(t,b)/Y(t)$ the current probability density functions of corresponding parameters in uninfected and infected populations accordingly; then

$$E_a(t) = \left(\int_A ax(t,a)da\right)\!\!\Big/X(t) = \int_A ap_1(t,a)da, \quad E_b(t) = \left(\int_B by(t,b)db\right)\!\!\Big/Y(t) = \int_B bp_2(t,b)db$$

are the current mean values of the parameters $a$ and $b$.

We suppose that the initial population sizes and initial distributions of the parameters are given:

$$x(0,a) = X_0 p_1(0,a), \quad y(0,b) = Y_0 p_2(0,b).$$

The particular form of system (A.2) is motivated by the models explored in the paper, which are special cases of (A.2).

Problem (A.2) can be reduced to the Cauchy problem for a certain system of ODE. Let

$$M_a(\lambda) = \int_A \exp(\lambda a)p_1(0,a)da,$$
(A.3)

$$M_b(\lambda) = \int_B \exp(\lambda b) p_2(0, b) db$$

be the mgfs (moment generating functions) of the initial probability densities $p_1(0,a)$ and $p_2(0,b)$.

Let us consider the system for auxiliary variables

$$dq^1_0(t)/dt = q^1_0(t) F_1(X^*, Y^*, E_1(t)), \quad q^1_1(0) = 1; \qquad (A.4)$$
$$dq^1_1(t)/dt = G_1(X^*, Y^*, E_1(t)), \quad q^1_1(0) = 0;$$
$$dq^2_0(t)/dt = q^2_0(t) F_2(X^*, Y^*, E_2(t)), \quad q^2_0(0) = 1;$$
$$dq^2_1(t)/dt = G_2(X^*, Y^*, E_2(t)), \quad q^2_1(0) = 0.$$

The functions $X^*(t)$, $Y^*(t)$ and $E_i(t)$ are defined by the formulas

$$E_1(t) = \frac{1}{M_a(q^1_1(t))} \left.\frac{dM_a(\lambda)}{d\lambda}\right|_{\lambda = q^1_1(t)}, \quad E_2(t) = \frac{1}{M_b(q^2_1(t))} \left.\frac{dM_b(\lambda)}{d\lambda}\right|_{\lambda = q^2_1(t)} \qquad (A.5)$$

$$X^*(t) = X(0) q^1_0(t) M_a(q^1_1(t)), \quad Y^*(t) = Y(0) q^2_0(t) M_b(q^2_1(t)).$$

**Theorem 1.** *Let Cauchy problem* (A.4)-(A.5) *have a unique global solution* $\{q^j_i(t)\}$ *at $t \in [0,T)$ where $0 \leq T \leq \infty$. Then the functions*

$$x(t,a) = x(0,a) q^1_0(t) \exp[a\, q^1_1(t)], \qquad (A.6)$$
$$y(t,b) = y(0,b) q^2_0(t) \exp[b\, q^2_1(t)],$$
$$X(t) = X(0) q^1_0(t) M_a(q^1_1(t)),$$
$$Y(t) = Y(0) q^2_0(t) M_b(q^2_1(t))$$

*satisfy system* (A.2) *at $t \in [0,T)$.*

*Conversely, if $x(t,a)$, $y(t,b)$, and $X(t) = \int_A x(t;a) da$, $Y(t) = \int_B y(t;b) db$ satisfy system* (A.2) *at $t \in [0,T)$, then Cauchy problem* (A.4) *has a global solution $\{q^j_i(t)\}$ at $t \in [0,T)$ and these functions can be written in the form* (A.6) *at $t \in [0,T)$.*

Proof.

Let $\{q^i_j(t)\}$ be a solution of Cauchy problem (A.4)-(A.5) at $t \in [0,T)$. According to system (A.4),

$$dx(t,a)/dt = x(0, a) \exp[a\, q^1_1(t)] \{dq^1_0(t)/dt + q^1_0(t)a\, dq^1_1(t)/dt\} =$$
$$x(t,a)\{F_1(X^*, Y^*, E_1(t)) + a\, G_1(X^*, Y^*, E_1(t))\}\,;$$
$$dy(t,b)/dt = y(0, b) \exp[b\, q^2_1(t)] \{dq^2_0(t)/dt + q^2_0(t)b\, dq^2_1(t)/dt\} =$$
$$y(t,b)\{F_2(X^*, Y^*, E_2(t)) + b\, G_2(X^*, Y^*, E_2(t))\}.$$

Next,

$$\int_A x(t;a)da = \int_A x(0,a) \exp[a\, q^1_1(t)]\, da = X(0)\, M_a(q^1_1(t));$$
$$\int_B y(t;b)db = \int_B y(0,b) \exp[b\, q^2_1(t)]\, db = Y(0)\, M_b(q^2_1(t)).$$

Hence, the functions defined by (A.6), are connected by the relations

$$X(t) = \int_A x(t,a)da,\ Y(t) = \int_B y(t;b)db.$$

To complete the proof of the first part of the theorem, we have to determine the current parameter distributions, $p_1(t,a)$ and $p_2(t,b)$, and show that $E_i(t)$ defined by (A.5) are equal to the current mean values of the parameters, i.e., $E_a(t) = \int_A a\, p_1(t,a)da = E_1(t)$ and $E_b(t) = \int_B b\, p_2(t,b)db = E_2(t)$. For this, we need the following

**Proposition 1**. *Let the Cauchy problem* (A.4)-(A.5) *have a unique solution* $\{q^i_j(t)\}$ *at* $t \in [0,T)$. *Then, for all* $t \in [0,T)$,

1) *the current parameter distributions,* $p_1(t,a)$ *and* $p_2(t,b)$ *are equal to*

$$p_1(t,a) = p_1(0,a)\, \exp[a\, q^1_1(t)] / M_a(q^1_1(t)), \qquad (A.7)$$

$$p_2(t,b) = p_2(0,b) \exp[b\, q^2{}_1(t)] / M_b(q^2{}_1(t));$$

2) *the moment generation functions of* $p_1(t,a)$ *and* $p_2(t,b)$ *are equal to*

$$M_a(t;\lambda) = M_a(\lambda + q^1{}_1(t)) / M_a(q^1{}_1(t)), \qquad (A.8)$$
$$M_b(t;\lambda) = M_b(\lambda + q^2{}_1(t)) / M_b(q^2{}_1(t))$$

*where* $M_a(\lambda)$, $M_b(\lambda)$ *are the mgf of initial distributions* (A.3).

Proof.

By definition, $p_1(t,a) = x(t,a)/X(t)$, $p_2(t,b) = y(t,b)/Y(t)$. According to (A.6),

$$p_1(t,a) = x(0,a)\, q^1{}_0(t) \exp[a\, q^1{}_1(t)] / X(t).$$

Further, $X(t) = X(0)\, q^1{}_0(t)\, M_a(q^1{}_1(t))$ by (A.5), so

$$p_1(t,a) = p_1(0,a) \exp[a\, q^1{}_1(t)] / M_a(q^1{}_1(t)).$$

Similarly, $p_2(t,b) = p_2(0,b) \exp[b\, q^2{}_1(t)] / M_b(q^2{}_1(t))$. Next,

$$M_a(t;\lambda) = \int_A \exp(\lambda a)\, p_1(t,a)\, da = \{\int_A \exp[(\lambda + q^1{}_1(t))a]\, p_1(0,a)\, da\} / M_a(q^1{}_1(t)) =$$
$$M_a(\lambda + q^1{}_1(t)) / M_a(q^1{}_1(t));$$

Similarly, $M_b(t;\lambda) = M_b(\lambda + q^2{}_1(t)) / M_b(q^2{}_1(t))$, as desired.

Now we are able to complete the proof of part 1 of Theorem:

$$E_a(t) = \int_A a\, p_1(t,a)\, da = \partial M_a(t;\lambda)/\partial\lambda \big|_{\lambda=0} = [\partial M_a(\lambda + q^1{}_1(t))/\partial\lambda \big|_{\lambda=0}] / M_a(q^1{}_1(t)),$$

and the last expression coincides with the formula (A.5) for $E_1(t)$.

Similarly, $E_b(t) = \partial M_b(t;\lambda)/\partial\lambda \big|_{\lambda=0} = [\partial M_b(\lambda + q^2{}_1(t))/\partial\lambda \big|_{\lambda=0}] / M_b(q^2{}_1(t))$.

Let us prove now part 2 of Theorem; let $x(t, a)$, $y(t,b)$, and $X(t) = \int x(t;a)da$, $Y(t) = \int y(t;b)db$ satisfy system (A.2) at $t \in [0,T)$, so that

$$\frac{\partial x(t,a)}{\partial t} = x(t,a)[F_1(X,Y,E_b(t)) + aG_1(X,Y,E_b(t))],$$

$$\frac{\partial y(t,b)}{\partial t} = y(t,b)[F_2(X,Y,E_a(t)) + bG_2(X,Y,E_a(t))],$$

where

$$E_a(t) = \left(\int_A ax(t,a)da\right)/X(t), \quad E_b(t) = \left(\int_B by(t,b)db\right)/Y(t).$$

Let us temporarily define the functions

$$q^1{}_0(t) = \exp[\int_0^t F_1(X(s),Y(s),E_b(s))ds], \quad q^1{}_1(t) = \int_0^t G_1(X(s),Y(s),E_b(s))\,ds,$$

$$q^2{}_0(t) = \exp[\int_0^t F_2(X(s),Y(s),E_a(s))ds], \quad q^2{}_1(t) = \int_0^t G_2(X(s),Y(s),E_a(s))\,ds,$$

Then

$$dx(t, a)/x(t,a) = dq^1{}_0(t)/q^1{}_0(t) + adq^1{}_1(t),$$
$$dy(t, b)/y(t,b) = dq^2{}_0(t)/q^2{}_0(t) + bdq^2{}_1(t);$$

hence

$$\ln x(t,a) = \ln q^1{}_0(t) + aq^1{}_1(t) + C^1$$
$$\ln y(t, b) = \ln q^2{}_0(t) + bq^2{}_1(t) + C^2$$

where $C^j$ do not depend on $t$.

Taking $C^1 = \ln x(0,a)$, $C^2 = \ln y(0,b)$ we obtain that

$$x(t,a) = x(0,a)\, q^1_0(t)\, \exp[b q^1_1(t)],$$
$$y(t,b) = y(0,b)\, q^2_0(t)\, \exp[b q^2_1(t)].$$

Hence

$$X(t) \equiv \int x(t,a)\,da = \int x(0,a)\, q^1_0(t)\, \exp[b q^1_1(t)]\, da = X(0)\, q^1_0(t)\, M_a(q^1_1(t)),$$

$$Y(t) \equiv \int y(t,b)\,db = \int y(0,b)\, q^2_0(t)\, \exp[b q^2_1(t)]\,db = Y(0)\, q^2_0(t)\, M_b(p^2_1(t)) \quad \text{at } t \in [0,T).$$

From the definition, $\{q^i_j(t)\}$ is a solution of Cauchy problem (A.4)-(A.5) for $t \in [0,T)$.

The proof is completed.

**Corollary.** System (A.2) is equivalent to the following non-autonomous system of ODEs:

$$\frac{dX}{dt} = X[F_1(X,Y,E_b(t)) + E_a(t) G_1(X,Y,E_b(t))],$$
$$\frac{dY}{dt} = Y[F_2(X,Y,E_a(t)) + E_b(t) G_2(X,Y,E_a(t))], \quad (A.9)$$
$$X(0) = X_0,\ Y(0) = Y_0$$

where the mean parameter values are

$$E_a(t) = \frac{1}{M_a(q_1(t))} \left.\frac{dM_a(\lambda)}{d\lambda}\right|_{\lambda=q_1(t)}, \quad E_b(t) = \frac{1}{M_b(q_2(t))} \left.\frac{dM_b(\lambda)}{d\lambda}\right|_{\lambda=q_2(t)} \quad (A.10)$$

where $M_a(\lambda)$ and $M_b(\lambda)$ are the mgf of the given initial pdfs $p_1(0,a)$ and $p_2(0,b)$, and the auxiliary variables $q_1(t)$, $q_2(t)$ can be found from the system

$$\frac{dq_1}{dt} = G_1(X,Y,E_a(t)), \quad \frac{dq_2}{dt} = G_2(X,Y,E_b(t)), \quad q_1(0) = 0, q_2(0) = 0. \qquad (A.11)$$

Proof.

It is enough to prove that system (A.9)-(A.11) is equivalent to (A.4)-(A.5).

Let Cauchy problem (A.4)-(A.5) have a unique global solution $\{q^j_i(t)\}$ at $t \in [0,T)$. Then, according to Theorem 1, functions (A.6) solve (A.2); it is easy to check that $X(t) = X(0)\, q^1{}_0(t)\, M_a(q^1{}_1(t))$, $Y(t) = Y(0)\, q^2{}_0(t)\, M_b(q^2{}_1(t))$ solve (A.9) and the definitions of the auxiliary variables and mean values of the parameters coincide with (A.10)-(A.11).

Conversely, let $X(t)$, $Y(t)$, $q_1(t)$, $q_2(t)$ solve (A.9)-(A.11) at $t \in [0,T)$. Let $q^1{}_1(t) = q_1(t)$, $q^2{}_1(t) = q_2(t)$ and

$$q^1{}_0(t) = \exp\left[\int_0^t F_1(X(s),Y(s),E_b(s))ds\right], \quad q^2{}_0(t) = \exp\left[\int_0^t F_2(X(s),Y(s),E_a(s))ds\right].$$

Then $\{q^j_i(t)\}$ is a solution of Cauchy problem (A.4)-(A.5); according to Proposition 1, the mgfs of the current distributions are given by (A.8) and mean parameter values (A.10) coincide with that defined by formulas (A.5). Q.E.D.

Proposition 1 allows one to determine the evolution of any initial parameter distribution with known mgf.

**Example**. Let parameter $a$ have the $\Gamma$-distribution at the initial instant with coefficients $s$, $\eta$, $k$:

$$P(0,a) = s^k (a - \eta)^{k-1} \exp[-(a - \eta)s] / \Gamma(k),$$

with the mean $\mu = \eta + k/s$, the variance $\sigma^2 = k/s^2$ and the mgf $M_a(\lambda) = \exp(\lambda\eta)/(1-\lambda/s)^k$. Then at any time moment $t < T$

$$M_a(t; \lambda) = M_a(\lambda + q_1(t)) / M_a(q_1(t)) = \exp(\lambda\eta)/(1-\lambda/(s - q_1(t)))^k,$$

i.e., the parameter $a$ has again the $\Gamma$-distribution at $t$ moment with coefficients $s-q_1(t)$, $\eta$, $k$. Next,

$$\partial M(\lambda+q_1(t))/\partial\lambda\,|_{\lambda=0}\,/\,M(q_1(t)) = k/(s-q_1(t)) + \eta.$$

Hence, the current mean of the parameter is

$$E_a(t) = \frac{k}{s-q_1(t)} + \eta \qquad (A.12)$$

where the auxiliary variable $q_1(t)$ is defined by (A.11).

As a particular inhomogeneous community model let us consider system (7):

$$\frac{\partial x(t,\beta)}{\partial t} = x(t,\beta)[1-(X(t)+Y(t))] - \frac{\beta x(t,\beta)Y(t)}{X(t)+Y(t)},$$

$$\frac{\partial y(t,\delta)}{\partial t} = \gamma y(t,\delta)[1-(X(t)+Y(t))] + \frac{E_\beta(t)X(t)y(t,\delta)}{X(t)+Y(t)} - \delta y(t,\delta).$$

Suppose that, at the initial moment, both parameters $\beta$ and $\delta$ have the $\Gamma$-distribution with coefficients $s_1$, $\eta_1$, $k_1$ and $s_2$, $\eta_2$, $k_2$, respectively. Then the current mean parameter values are given by (A.12) and, finally, system (7)-(8) is reduced to the following system of ODEs:

$$\frac{dq_1}{dt} = -\frac{Y}{X+Y}, \quad \frac{dq_2}{dt} = -1,$$

$$\frac{dX}{dt} = X(1-(X+Y)) - \left(\frac{k_1}{s_1-q_1(t)} + \eta_1\right)\frac{XY}{X+Y},$$

$$\frac{dY}{dt} = \gamma Y(1-(X+Y)) + \left(\frac{k_1}{s_1-q_1(t)} + \eta_1\right)\frac{XY}{X+Y} - \left(\frac{k_2}{s_2-q_2(t)} + \eta_2\right)Y,$$

$$X(0) = X_0, Y(0) = Y_0, \quad q_1(0) = 0, q_2(0) = 0.$$